\documentclass[aps,prl,reprint,amsmath,amssymb,superscriptaddress,nobibnotes,showpacs]{revtex4-1}

\usepackage{graphicx}
\usepackage{dcolumn}
\usepackage{bm}
\usepackage{amsmath}
\usepackage{amssymb,amsthm}
\usepackage{color}
\usepackage[sans]{dsfont}

\definecolor{nblue}{rgb}{0.2,0.2,0.7}
\definecolor{ngreen}{rgb}{0.2,0.6,0.2}
\definecolor{nred}{rgb}{0.7,0.2,0.2}
\definecolor{nblack}{rgb}{0,0,0}

\newcommand{\tr}{\text{tr}}

\newcommand{\Id}{\openone}

\def\tr{\mbox{tr}}
\def\bea{\begin{eqnarray}}
\def\eea{\end{eqnarray}}

\begin{document}

\title{Realizing Physical Approximation of the Partial Transpose}


\author{Hyang-Tag Lim}
\email{forestht@gmail.com}
\affiliation{Department of Physics, Pohang University of Science and Technology (POSTECH), Pohang, 790-784, Korea}

\author{Yong-Su Kim}
\affiliation{Department of Physics, Pohang University of Science and Technology (POSTECH), Pohang, 790-784, Korea}

\author{Young-Sik Ra}
\affiliation{Department of Physics, Pohang University of Science and Technology (POSTECH), Pohang, 790-784, Korea}

\author{Joonwoo Bae}
\email{bae.joonwoo@gmail.com}
\affiliation{School of Computational Sciences, Korea Institute for Advanced Study, Seoul, 130-012, Korea}

\author{Yoon-Ho Kim}
\email{yoonho72@gmail.com}
\affiliation{Department of Physics, Pohang University of Science and Technology (POSTECH), Pohang, 790-784, Korea}

\date{\today}


\begin{abstract}
The partial transpose by which a subsystem's quantum state is solely transposed is of unique importance in quantum information processing from both fundamental and practical point of view. In this work, we present a practical scheme to realize a physical approximation to the partial transpose using local measurements on individual quantum systems and classical communication. We then report its linear optical realization and show that the scheme works with no dependence on local basis of given quantum states. A proof-of-principle demonstration of entanglement detection using the physical approximation of the partial transpose is also reported.
\end{abstract}


\maketitle


\textit{Introduction.--} One of the major challenges in quantum information processing is to characterize capabilities in manipulating quantum states for information tasks. The postulates of quantum theory in fact dictate which operations are allowed to do, or not. The transpose is one of the non-physical cases, and the reason lies at the most fundamental level of quantum theory: a combination of unitary and anti-unitary transformations is not a transformation that preserves physical symmetries \cite{ref:wigner}. An example of the case would be with the identity and the transpose,  denoted by $\openone \otimes T$ and called the partial transpose  by which only a subsystem's state is transposed. As pointed out by Peres, the partial transpose (PT) operation which is of  fundamental interest, albeit non-physical, has an important application in detecting entanglement of quantum states \cite{ref:Peres}.

Entanglement is generally a resource for quantum information applications \cite{ref:qent}. Indeed, all entangled states are useful for certain information tasks, e.g. quantum teleportation \cite{ref:masanes}. Given composite quantum systems for such applications, it is then required to determine if they are in entangled states rather than to identify their quantum states. This naturally defines the so-called \textit{direct detection} of entanglement \cite{ref:toth}. Due to the impossibility of directly applying the PT in experiment, much effort has been devoted to developing an experimentally feasible formalism, entanglement witnesses that are based on \emph{local observables rather than operations} \cite{ref:lewen}.

In Ref.~\cite{ref:HorodeckiEkert}, Horodecki and Ekert proposed a method called structural physical approximation (SPA) by which non-physical operations that can detect entangled states such as the PT can be systematically approximated by physical operations. Moreover, SPAs to these operations can be factorized into local operations and classical communication (LOCC) \cite{ref:spalocc}. Being based on applications of \emph{operations rather than observables}, entanglement detection using SPAs works with no dependence on local basis of given quantum states \cite{ref:note}. Using the SPA to PT (SPA-PT), therefore, all entangled states of two qubits as well as other useful entangled states in high-dimensions can be detected \cite{ref:Horodecki}. On the practical side, it was recently suggested in Ref.~\cite{ref:Korbicz} that SPAs to optimal positive maps can be in general replaced by quantum channels of measurement followed by preparation of quantum states. This in fact significantly improves the experimental feasibility of the direct detection of entanglement using SPAs within present-day technology. It was also shown that the SPA-PT can be done in the same way \cite{ref:Korbicz, ref:fiu}.

Therefore, for an \emph{operation-based approach} to the direct detection of entanglement \cite{ref:HorodeckiEkert}, it is of utmost importance to devise a practical SPA-PT scheme and demonstrate its feasibility towards entanglement detection. We also emphasize from the fundamental point of view that, realizing non-physical operations (i.e. not allowed in quantum theory) in their approximate and optimal forms would characterize and confirm how far one can go in manipulating quantum states for information tasks within the fundamental limit. So far, apart from the experimental feasibility \emph{per se}, little is known about how to carry out the SPA-PT in practice with minimal experimental resources.

In this Letter, we provide and demonstrate a practical scheme to realize the SPA-PT for two-qubit states, based solely on local measurements and classical communication. The experimental demonstration of the SPA-PT scheme is performed in photonic systems, i.e., using single-photon polarization qubits and linear optical devices. The results show that the proposed scheme works equally well for all Bell-states, indicating no dependence on local basis. Finally, a proof-of-principle demonstration of entanglement detection using the SPA-PT is shown.


\textit{Scheme.--} Let us begin by describing the theoretical scheme to realize the SPA-PT. The central idea of the SPA to a linear map $\Lambda$ lies on the fact that by admixing with the depolarization, $D [\rho] = I_d /d$ where $I_d$ is the identity matrix in a $d$ dimension, the map $\Lambda$ can be transformed to a completely positive map $\widetilde{\Lambda}$. The SPAed map $\widetilde{\Lambda} = (1-p) \Lambda +p D$ with minimum $p\geq 0$ represents a physical operation \cite{ref:phorodecki}. For linear maps $\openone \otimes \Lambda$ that can detect entangled states, the SPA then works as $\widetilde{\openone \otimes \Lambda} = (1-p) \openone \otimes \Lambda + p D\otimes D$ with minimum $p\geq0$ \cite{ref:HorodeckiEkert} and moreover can be factorized into a form of LOCC \cite{ref:spalocc}. Having collected all these facts and applying to the case of the PT, one can derive the following decomposition for the SPA-PT for a two-qubit state $\rho_{AB}$,
\bea (\widetilde{\openone\otimes T})[\rho_{AB}] & = &\frac{1}{3} (\openone\otimes \widetilde{T}) [\rho_{AB}]+ \frac{2}{3} (\widetilde{\Theta} \otimes D)[\rho_{AB}],~~ \label{eq:PTlocc} \eea where $\widetilde{T}$ and $\widetilde{\Theta}$ denote SPAs to the transpose and to the inversion, respectively, where the inversion $\Theta[\rho] = -\rho$.

The SPAed transpose $\widetilde{T}$, as shown in Ref.~\cite{ref:Korbicz},  corresponds to a channel based on measurement and preparation of quantum states  and  it can be written as \cite{ref:Lim}
$ \widetilde{T}\left[ \rho  \right] = \sum\limits_{k = 1}^4 {{\rm{tr}}\left[ M_{k} \rho \right]\left| {v_k } \right\rangle \left\langle {v_k } \right|}$ for a state $\rho$, where
\bea \left| {v_1 } \right\rangle  & \propto & {\left| 0 \right\rangle  + \frac{{ie^{i\pi 2/3} }}{{i + e^{ - i\pi 2/3} }}\left| 1 \right\rangle }, ~
\left| {v_2 } \right\rangle  \propto  {\left| 0 \right\rangle  - \frac{{ie^{i\pi 2/3} }}{{i - e^{ - i\pi 2/3} }}\left| 1 \right\rangle}, \nonumber\\
\left| {v_3 } \right\rangle  & \propto &  {\left| 0 \right\rangle  + \frac{{ie^{i\pi 2/3} }}{{i - e^{ - i\pi 2/3} }}\left| 1 \right\rangle }, ~
\left| {v_4 } \right\rangle  \propto {\left| 0 \right\rangle  - \frac{{ie^{i\pi 2/3} }}{{i + e^{ - i\pi 2/3} }}\left| 1 \right\rangle },\nonumber \eea
and $\{ M_{k} = |v_{k}^{*}\rangle \langle v_{k}^{*} |/2 \}_{k=1}^{4}$ is a complete measurement. The SPAed inversion  $\widetilde{\Theta}$ also corresponds to a channel based on measurement followed by state preparation, and can be expressed using the Pauli matrix $\sigma_y$ as, $\widetilde{\Theta} [\cdot] = \sigma_{y} \widetilde{T}[\cdot] \sigma_{y}$.


\begin{figure}[t]
\includegraphics[width=3.4in]{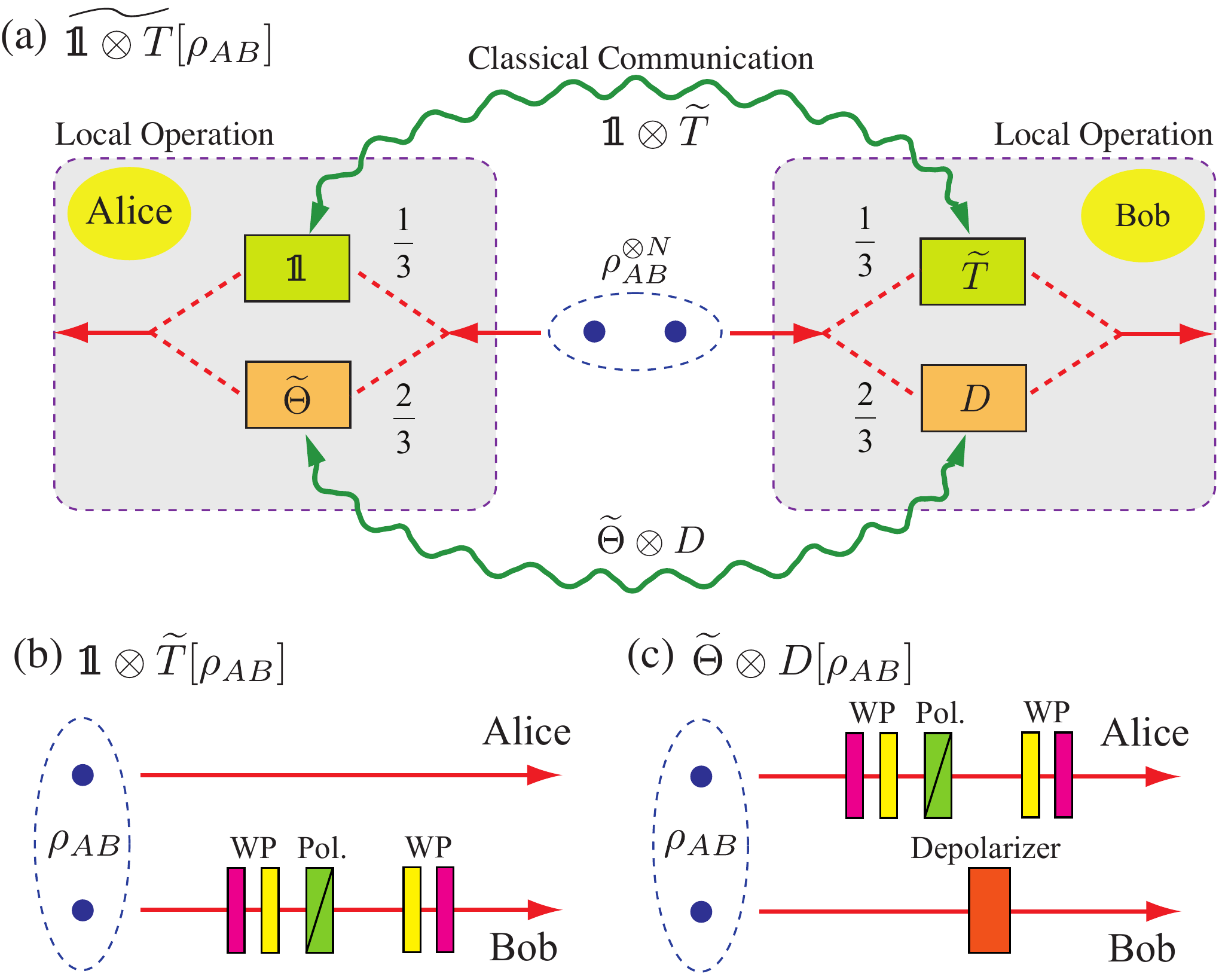}
\caption{The SPA-PT, $\widetilde{\Id \otimes T}$, for the input state $\rho_{AB}$  is accomplished by applying $\Id \otimes \widetilde{T}$ and $\widetilde{\Theta} \otimes D$ with probabilities $1/3$ and $2/3$, respectively, shown in (a). Combinations of waveplates (WP) and polarizers (Pol.) implement measurement and preparation of quantum states, shown in (b) and (c). See text for details}\label{fig:setup}
\end{figure}

\textit{Realization.--} The SPA-PT in eq. (\ref{eq:PTlocc}) can be implemented by applying $\openone \otimes \widetilde{T}$ and $\widetilde{\Theta} \otimes D$ with probabilities $1/3$ and $2/3$ respectively, see Fig. \ref{fig:setup} (a). The local operations are actually (single-copy) \textit{measurements} followed by \textit{state preparation}, see Fig. \ref{fig:setup} (b) and (c). Measurement $M_{k}$, i.e. on the basis $|v_{k}^{*}\rangle = U_{k} |0\rangle$, is performed by unitary transformation $U_k$ and measurement in the computational basis. The preparation step can be done by transforming a state collapsed by measurement, to a corresponding one using optical elements \cite{ref:Lim}. For photonic polarization qubits, $|0\rangle = |H\rangle$ and  $|1\rangle = |V\rangle$, waveplates and a polarizer are optical elements to perform unitary transformations and the measurement in the computational basis, respectively. If a single-photon is found after passing through a set of waveplates for $U_k$ and a polarizer aligned for measurement in $|0\rangle\langle 0|$, it would mean that the qubit has collapsed to the state $|0\rangle$ due to the measurement. The collapsed state $|0\rangle$ would then be used for state preparation according to $\widetilde{T}$ or $\widetilde{\Theta}$, see Fig. \ref{fig:setup} (b) and (c). The depolarization $D$ can be performed by random applications of Pauli matrices, $D\left[\cdot \right] = 1/4\sum_{i = 0,x,y,z} {\sigma _i [\cdot] \sigma _i }$, with each $\sigma_i$ implemented by waveplates.

The details for the experimental realization are the following. A two-qubit state is prepared using the spontaneous parametric down-conversion (SPDC) process. A 6 mm thick type-I BBO crystal is used in the frequency-degenerate, non-collinear phase matching condition. The BBO crystal was pumped by a 405 nm diode laser beam (100 mW) and the SPDC photon pairs were centered at 810 nm. All four two-qubit Bell states (i.e., two-photon polarization Bell states) are then prepared with quantum interferometry \cite{ref:Ou}. We observed around 900 Hz coincidence counting rate using 5 nm full width at half maximum bandpass filters.

\begin{figure}[t]
\includegraphics[width=3.4in]{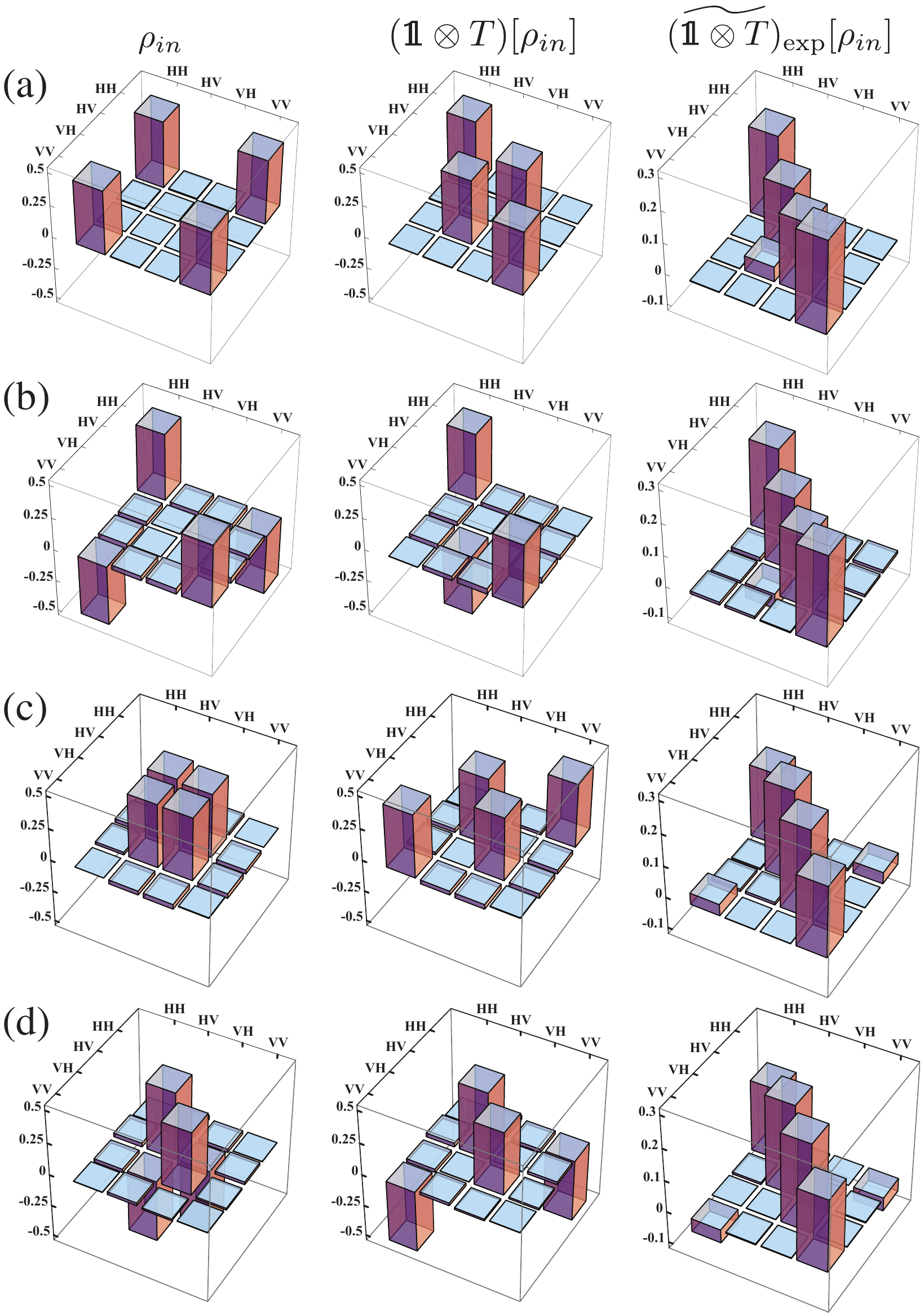}
\caption{The 1st column shows QST of the four Bell states prepared in experiment $\rho_{in}$: (a) $|\phi^{+}\rangle$, (b) $|\phi^{-}\rangle$, (c) $|\psi^{+}\rangle$, and (d) $|\psi^{-}\rangle$). The resulting states after applying the SPA-PT are identified by QST, shown in the 3rd column. Note that only real parts of the density matrices are  shown as the imaginary parts are almost zero. These can be compared with the ideal case of the PT, shown in the 2nd column.}
\label{fig:QST}
\end{figure}

Let us now demonstrate the SPA-PT scheme for Bell states $|\phi^{\pm} \rangle = (|00\rangle \pm |11\rangle)/\sqrt{2}$, $|\psi^{\pm}\rangle = (|01\rangle \pm |10\rangle)/\sqrt{2}$. These states are particularly chosen to show that the performance depends largely on correlations existing in quantum states but not their local basis. This is indeed an important feature toward efficient detection of entanglement \cite{ref:note}. In experiment, the initially prepared Bell states and the resulting states after the SPA-PT are identified using quantum state tomography (QST) \cite{ref:qst}. The experimental results are shown in Fig.~\ref{fig:QST}.

To quantify the performance, we compute the Uhlmann's fidelity $F\left( {\rho,\sigma } \right) = \left[ {\rm{tr} {\sqrt {\sqrt \rho  \sigma \sqrt \rho  } } } \right]^2$ between two states, one from the experimental realization $\widetilde{(\Id \otimes T)}_{\rm{exp}} $ and the other from the ideal one $\widetilde{(\Id \otimes T)}$ \cite{ref:fidelity}. For the states $|\phi^{\pm}\rangle$, $|\psi^{-}\rangle$, our experiment shows $F=0.999$, and for $|\psi^{+}\rangle$, $F=0.998$. These results show that for the proposed SPA-PT scheme only based on LOCC, the experimental realization works faithfully without a dependence on local basis.

\textit{Application.--} So far, a scheme to realize the SPA-PT based on single-copy measurement and preparation of quantum states is proposed and its experimental realization is shown. An important application of the SPA-PT, as it was originally proposed for, is entanglement detection of unknown quantum states \cite{ref:HorodeckiEkert}. Applying the PT to a two-qubit state, a negative eigenvalue is a sufficient condition to conclude that the state is entangled. When applying the SPA-PT, the condition would be an eigenvalue smaller than the portion of admixed noise. It is an eigenvalue found to be smaller than $2/9$ for two-qubit states \cite{ref:HorodeckiEkert}. Thus, once the SPA-PT is applied to unknown two-qubit states, to determine if they are entangled requires to estimate minimum eigenvalues of resulting quantum states.

It has been known that a general method to obtain the minimum eigenvalue of unknown quantum states requires collective measurement for which one should be able to store quantum states for a while \cite{ref:state-esti}. Since the SPA-PT is now performed by local measurements, the spectrum estimation no longer requires collective measurement on resulting quantum states. This simplifies experimental resources and thus hugely improves the practical feasibility \cite{ref:Korbicz}. Estimating eigenvalues of resulting states after SPA-PT then defines a classical optimization problem over measurement outcomes.

Here, the goal is a proof-of-principle demonstration for detecting entanglement (of unknown quantum states) from measurement outcomes of the SPA-PT only. To this end, we consider a ``brute-force" approach in the following, in the sense that the cost such as measurement settings is not optimized in terms of its efficiency at this stage. This means that all measurement outcomes are collected from $\widetilde{T}$, $\widetilde{\Theta}$ and $D$ and applied to determining if given quantum states are entangled or separable. For the identity operation in $\openone\otimes \widetilde{T}$, any measurement in a tomographically complete basis $\{|t_i\rangle \}_{i=1}^{4}$ is applied. In experiment, measurements of $\Id \otimes \widetilde{T}$ and, similarly, measurements of $\widetilde{\Theta} \otimes D$ are repeated for the two-qubit state $\rho_{AB}$. The probabilities are then obtained from these measurement outcomes: $p_{ij} = \tr\left[ {\rho_{AB} |t_{i}\rangle \langle t_{i}| \otimes M_j } \right]$, $q_k  = \tr\left[ {\rho_{AB} M_k  \otimes |0\rangle\langle 0 |} \right]$, and $r_k  = \tr\left[ {\rho_{AB} M_k  \otimes |1\rangle\langle 1|} \right]$ where the states $\{|t_{i}\rangle \}_{i=1}^{4}$ are chosen, for convenience, from $\left\{ |0\rangle, |1\rangle, (|0\rangle + |1\rangle)/\sqrt{2},  (|0\rangle +i |1\rangle)/\sqrt{2} \right\}$. From these probabilities, it is possible to reconstruct an operator after the SPA-PT in eq.~(\ref{eq:PTlocc}) and determine the eigenvalues using determinant. The details are shown in the Appendix.

\begin{figure}[t]
\includegraphics[width=3.2in]{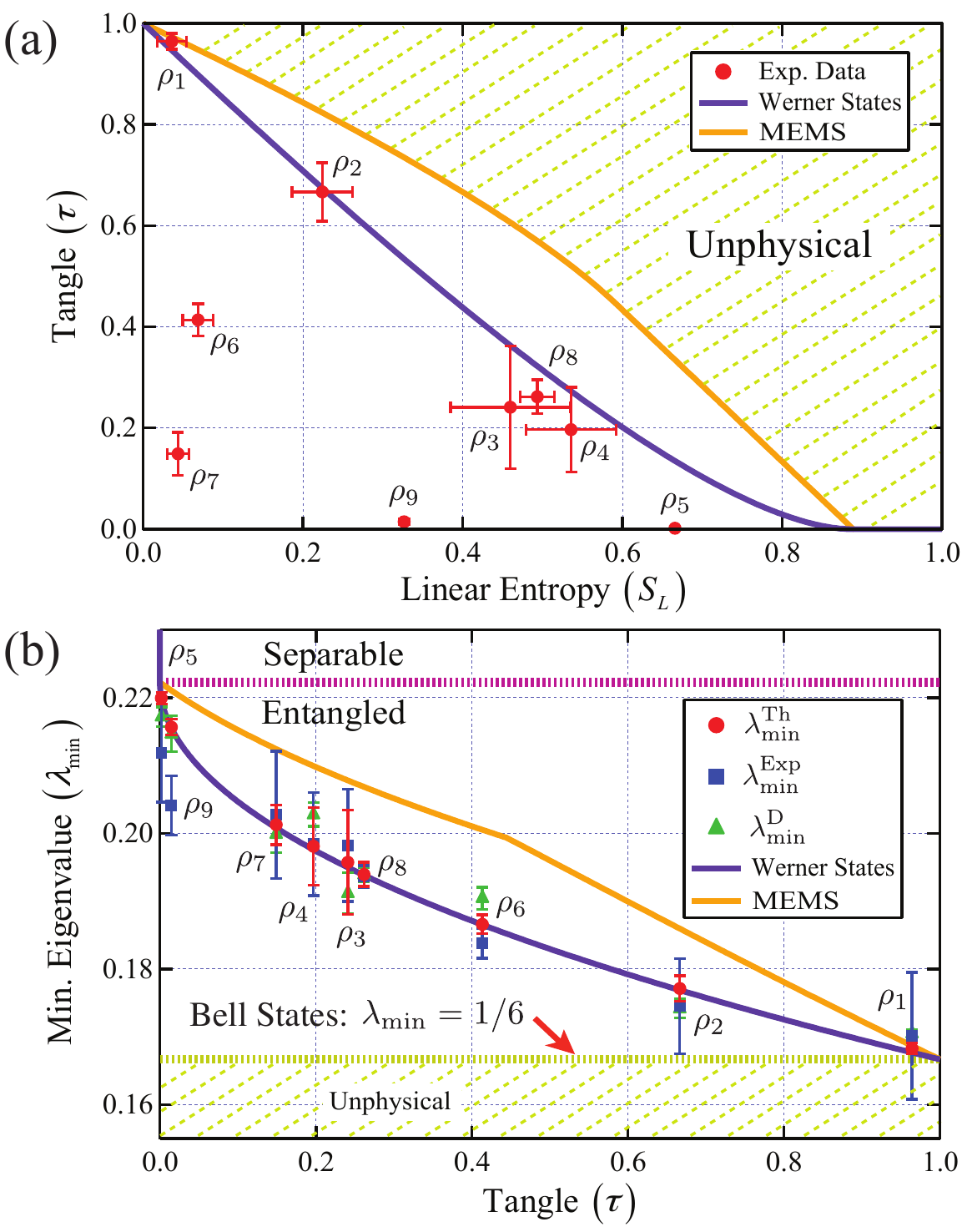}
\caption{ (a) Nine quantum states $\{\rho_k\}_{k=1}^{9}$ are shown in the tangle ($\tau$) - linear entropy ($S_L$) plane. (b) Minimum eigenvalues of resulting states after SPA-PT are shown. The error bars in (b) with respect to $\tau$ can be found in (a). Negative eigenvalues of two-qubit states after the PT is in the range $[-1/2,0]$, which is $[1/6,2/9]$ in the case of the SPA-PT. The Werner state $\rho _{\rm{W}}  = pI \otimes I/4 + \left( {1 - p} \right)\left| {\psi^-} \right\rangle \left\langle {\psi ^ -  } \right|$ and the Maximally entangled mixed states $\rho_{MEMS} = f(p)(|00\rangle \langle 00| + |11\rangle \langle 11|) + p(|00\rangle \langle 11| + |11\rangle \langle 00|)/2 + (1-2f(p))|01\rangle\langle 01|$ where $f(p)$ is $p/2$ for $p\geq2/3$ and $1/3$ for $p < 2/3$ \cite{ref:Munro}. }\label{fig:graph}
\end{figure}

The above ``brute-force'' method is applied to demonstrate entanglement detection of two-qubit states in the following form: $\rho(p,\alpha)  = \left( {1 - p} \right)\left| \psi  \right\rangle \left\langle \psi  \right| + p\left| {\psi ^ \bot  } \right\rangle \left\langle {\psi ^ \bot  } \right|$ where $|\psi\rangle = \alpha|01\rangle -\sqrt{1-|\alpha|^{2}} |10\rangle$. We generated nine different quantum states $\rho_{k} (p,\alpha)$ (for $k=1,\cdots,9$) and identified them using QST: $(p,\alpha)=\{(0,0.71)$, $(0.12, 0.71)$, $(0.25, 0.71)$, $(0.3,0.71)$, $(0.51, 0.71)$, $(0, 0.92)$, $(0, 0.97)$, $(0.37, 0.86)$, $(0.42, 0.92)\}$. Note that these $(p,\alpha)$ are obtained as the average of the data. In Fig.~\ref{fig:graph}(a), these states $\{ \rho_{k} \}_{k=1}^{9}$ are shown in terms of the linear entropy $S_{L}$ (as a measure of mixedness) and the tangle $\tau$ as an entanglement measure \cite{ref:wootters}.

For those quantum states $\{\rho_k \}_{k=1}^{9}$, we compare three cases of entanglement detection: a theoretical prediction, experimental results by realizing the SPA-PT, and the ``brute-force'' approach. Therefore, the minimum eigenvalue denoted by $\lambda_{\mathrm{min}}$ is obtained for each case, as follows: $\lambda_{ \mathrm{min} }^{ \rm{Th}}$ refers to the one from the ideal (theoretical) SPA-PT of the input state $\rho_k$, $\lambda_{ \mathrm{min}}^{ \rm{Exp}}$ is computed from the QST of the resulting state after the SPA-PT in experiment, and $\lambda_{\mathrm{min}}^{\rm{D}}$ is from the above-mentioned ``brute-force'' method. In Fig.~\ref{fig:graph} (b), all these are compared in terms of the minimum eigenvalues and the tangle $\tau$. We have also performed these for four Bell states and the results show that all minimum eigenvalues are indeed smaller than $2/9$, see Table~\ref{table:lambda}. With these extensive examples, we have verified experimentally that a method for entanglement detection using the SPA-PT does largely depend on correlations existing in quantum states.

\textit{Conclusion.--} We have provided a practical scheme to realize an approximate PT via the SPA and shown its linear optical realization for two-qubit states of photonic systems. Contrast to entanglement witnesses, it is shown that the realization works without a dependence on local basis. Since the scheme is based on local measurements and classical communication, it can be applied to long-distance quantum information tasks. As the partial transpose is an instance in Ref.~\cite{ref:Korbicz}, the presented construction can be generalized to realizing other non-physical operations that can detect entanglement in high-dimensional and multipartite quantum systems.

For a number of entangled states, we have reported a proof-of-principle demonstration of entanglement detection using the SPA-PT. With further classical optimizations, applications of the SPAs would lead to efficient and practical methods of detecting entanglement with much dependence on correlations existing in quantum states rather than on local basis.

A useful application would be secure quantum communication where, an adversary having an ultimate power to control quantum channels may change local basis such that entanglement cannot be detected by an employed detection method, e.g. entanglement witnesses \cite{ref:entinf}. In various contexts of quantum information processing, entanglement detection is generally and often a basic task required for quantum information applications. Accordingly, it would be interesting to optimize the cost for the detection (such as measurement settings) with respect to the efficiency.

\begin{table}
\begin{center}
\begin{tabular}{ccccccccc}
\hline
\hline
&~~~~& $|\phi^{+}\rangle$ & ~~~~ & $|\phi^{-}\rangle$ &~ ~~~ & $|\psi^{+}\rangle$ & ~~~~ & $|\psi^{-} \rangle$  \\
\hline
$\lambda_{\mathrm{min}}^{\rm{Th}}$ && $0.169$ && $0.170$ && $0.168$ && $0.168$  \\
\hline
$\lambda_{\mathrm{min}}^{\rm{Exp}}$ && $0.169$ && $0.166$ && $0.171$ && $0.170$  \\
\hline
$\lambda_{\mathrm{min}}^{\rm{D}}$ && $0.174$ && $0.173$ && $0.166$ && $0.171$   \\
\hline
\hline
\end{tabular}
\end{center}
\caption{Minimum eigenvalues $\lambda_{\rm{min}}$ obtained using three different methods described in the text for four Bell states shown in Fig.~\ref{fig:QST}. In all cases, $\lambda_{\rm{min}}$ is much smaller than the threshold $2/9 \approx 0.222$ and, in fact, very close to the maximal entanglement case,  $1/6 \approx 0.167$.
}
\label{table:lambda}
\end{table}

This work was supported by the National Research Foundation of Korea (2009-0070668 and 2009-0084473) and the Ministry of Knowledge and Economy of Korea through the Ultrafast Quantum Beam Facility Program.



\section*{Appendix: Application to detecting entanglement}

The SPA-PT was originally proposed in Ref.~\cite{ref:HorodeckiEkert} to directly making use of the PT in experiment such that entanglement can be directly detected in experiment. Compared to entanglement witnesses, this method is closer to the detection based on so-called \textit{positive but not completely positive maps} \cite{ref:Horodecki}, in the sense that entangled states are detected by operations rather than expectations to some specified observables. To be explicit, entanglement witnesses for bipartite quantum systems are constructed as follows. Given a positive map $\Lambda$ that is not completely positive, let $S_{\Lambda} = \{ \rho ~\| ~(\openone\otimes \Lambda)[\rho] \ngeq 0 \}$, the set of entangled states detected by the map. In other words, there exist a set of projectors, $S_{Q} = \{ Q ~\| ~ \tr[Q (\openone\otimes \Lambda)[\rho]]< 0 \}$. Then, the witnesses are constructed as, $S_{W}= \{ W~\| ~ W = (\openone\otimes \Lambda^{*}) (Q), \forall Q\in S_{Q} \}$ where it holds that $\tr[(\openone\otimes \Lambda^{*}) (Q) \rho]  = \tr[Q (\openone\otimes \Lambda) (\rho)]$. Therefore, one can apply either the operation $\Lambda$ or its dual $S_{W}$ to detect entangled states $S_{\Lambda}$. The advantage of an operation-based method is that, contrast to the witnesses, the detection does not depend on local basis.

An operation-based entanglement detection, in particular the SPA-PT, in Ref.~\cite{ref:HorodeckiEkert} can be summarized as follows. Implementing the SPA-PT to a number of quantum states, the resulting states are stored in quantum memory. Then, for those states, optimal spectrum estimation is applied using collective measurement. While measurement outcomes of the collective measurement in Ref.~\cite{ref:state-esti} directly give the minimum eigenvalue that answers if given quantum states were entangled or separable, quantum memory required to apply the collective measurement is an experimentally challenging issue to date. It was shown in Ref.~\cite{ref:fiu} that the SPA-PT for two-qubit states can be done by measurement and preparation of quantum states. Recently in Ref.~\cite{ref:Korbicz} it was shown that the SPA-PT can generally be done within present-day technology, that is, via measurement followed quantum state preparation. Therefore, all these lead to defining an optimization over measurement outcomes that gives a minimum eigenvalue. These are explicitly constructed in eq. (\ref{eq:PTlocc}).

The goal is the proof-of-principle demonstration that the SPA-PT can be applied to entanglement detection. We provide a function by which a minimum eigenvalue can be computed only from measurement outcomes. This can thought of as a ``brute-force" approach to detect entanglement since all measurement outcomes are considered. Optimization of the number of measurement basis needed with respect to its efficiency in detecting entanglement, i.e. number of entangled states that are detected, is to be done with further investigation. Remind that, for a given state $\rho$, probabilities are obtained from measurements, $p_{ij} = \tr\left[ {\rho_{AB} |t_{i}\rangle \langle t_{i}| \otimes M_j } \right]$, $q_k  = \tr\left[ {\rho_{AB} M_k  \otimes |0\rangle\langle 0 |} \right]$, and $r_k  = \tr\left[ {\rho_{AB} M_k  \otimes |1\rangle\langle 1|} \right]$ where POVMs $\{M_{i}\}_{i=1}^{4}$ are from realization of the SPA-PT. Using all of the measurement outcomes collected, one can construct the following operator, a function of the measurement outcomes,
\bea \hat{F}(p_{ij},q_k ,r_k ) & = & \frac{1}{3} \sum_{i,j=1}^4 p_{ij} |t_i \rangle\langle t_i|  \otimes M_{j} \nonumber \\ &&+\frac{1}{3} \sum_{k=1}^4 M_{k} \otimes ({q_k} |0\rangle \langle0| + {r_k} |1\rangle \langle 1| ). \nonumber \eea Compute $\lambda = \min \{\kappa ~\|~ \det{(\hat{F} -\kappa I)} =0 \}$, i.e. the eigenvalue that determines if the state $\rho$ is entangled or not. If it is larger than $2/9$, the state is separable, otherwise, entangled. Note that this may not coincide to eigenvalues obtained after QST since that the usual QST method, e.g. maximum likelihood estimation, is not applied here. It remains open to minimize or characterize the number of measurement basis to detect specific instances of entangled states. This also devises the function of measurement outcomes in the above. In the above, it is demonstrated that with a number of measurement outcomes only and a function of them, it can be determined if given quantum states are entangled or not.


\begin{thebibliography}{99}
\bibitem{ref:note} For entanglement witnesses often and widely used for entanglement detection, the efficiency highly depends on local basis in which a quantum state is entangled. For instance, a witness that detects a single Bell state can never detect the other Bell states.
\end{thebibliography}
\end{document}